\documentclass[aps,onecolumn,showpacs,groupedaddress]{revtex4-1}
\usepackage{amssymb}
\usepackage{mathrsfs}
\usepackage{pstricks}
\usepackage{color}
\usepackage{graphicx}
\usepackage[fleqn]{amsmath}
\usepackage{slashed}
\usepackage{amsmath}
\usepackage{mathtools}
\usepackage{array}
\usepackage{tabularx}
\usepackage{physics}
\usepackage{slashed}
\usepackage[utf8]{inputenc}
\usepackage[T1]{fontenc}
\usepackage{float}

\begin{document}


\title{\bf Non-ideal quantum electrodynamics}


\author{P. R. S. Carvalho}
\email{prscarvalho@ufpi.edu.br}
\affiliation{\it Departamento de F\'\i sica, Universidade Federal do Piau\'\i, 64049-550, Teresina, PI, Brazil}





\begin{abstract}
In this Letter we present a field-theoretic formulation for describing non-ideal quantum electrodynamic effects. It generalizes its ideal counterpart and is valid in the non-ideal domain. We compute some non-ideal elementary processes both at leading order as well as up to next-to-leading order. Furthermore, we present some physical effect taken up to next-to-leading order that is not described by earlier similar but non-renormalizable theories defined recently in literature thus characterizing all them, at most, as effective ones. Their effectiveness allows us to obtain such an effect only at leading order. We recover all the corresponding ideal results when we take the limit $a\rightarrow 1$, where $a$ is a parameter that encodes the non-ideal effects. All the results show an interplay between non-ideal properties and fluctuations.
\end{abstract}


\maketitle


\section{Introduction}

\par The generalized non-ideal $a$-statistics introduced in Ref. \cite{ArxivNISFT} has achieved great success through the agreement between their theoretical predictions and experimental data \cite{ArxivNISFT} for some non-ideal low-energy systems undergoing continuous phase transitions. Non-ideal systems ($a\neq 1$) present defects, impurities and inhomogeneities as opposed to ideal ones which are perfect, pure and homogeneous ($a = 1$). The corresponding field-theoretic formulation of $a$-statistics is called non-ideal statistical field theory (NISFT) and is capable of furnishing non-ideal results beyond the leading order (LO) in the number of loops, namely up to next-to-leading order (NLO), which was not the case for other generalized statistics which were limited to display results just at LO \cite{ArxivNISFT}. When, for some physical quantity of interest, its one-loop order term vanishes, its leading order term is the two-loop order one if that term does not vanish and so on. The critical exponent $\eta$ for systems undergoing continuous phase transitions is a typical case, where its one-loop contribution vanishes \cite{Wilson197475}. In fact, results restricted only up to one-loop order means that the corresponding theory is not fundamental. Thus, it is, at most, an effective one and its effects are probed only up to one-loop order and not beyond that loop order. This is the importance of defining a theory whose predictions go beyond one-loop order. On the other hand, NISFT attains that aim \cite{ArxivNISFT}. When a theory attains the aforementioned aim, we say that it is renormalizable and the physical effects can now be taken at all length of scales \cite{WilsonSciAme} or loop orders. This is why renormalization plays a central role in determining which theory is consistent for studying many-particle systems phenomena at the deepest and fundamental level. 

\par Some examples of non-renormalizable of effective theories are \cite{Schwartz}: the Schrödinger equation, the Fermi theory for weak force, chiral perturbation theory and general relativity. The consistent renormalizable or fundamental theory corresponding to three of the four effective interactions aforementioned is the Standard Model of elementary particles and interactions, \emph{i. e.}, it encompasses the fundamental electromagnetic, weak and strong forces, respectively. The effective theories aforementioned provided results that were limited only up to the one-loop order \cite{Schwartz}. But while the referred fundamental renormalizable theories were not yet achieved, the corresponding inconsistent effective non-renormalizable theories were employed in computations even if their results were not valid for all scales of length or not attained loop orders beyond the one-loop level. Although some results reached a numerical magnitude of the order of the correct results (in the case of Fermi theory \cite{Peskin}) they were, at least, conceptually wrong. In fact, these results were written as just a one-loop term but the true effects are a result of a sum of many contributions, namely one-, two-, three-loop order contributions etc. Thus, the fundamental theories were not limited only to one-loop order. They contain one- and high loop order contributions and the inconsistent effective non-renormalizable theories were discarded as fundamental ones and were regarded, at most, as effective ones. We have to show that the theory presented in this Letter can describe non-ideal quantum electromagnetic phenomena up to NLO as opposed to some non-renormalizable or effective theories defined recently in the literature, once the latter furnish the corresponding results limited only to LO.

\section{$S_{a}$-matrix}\label{Sa-matrix}

\par We consider the ideal bare Lagrangian density of quantum electrodynamics (QED) given by \cite{Peskin}
\begin{eqnarray}\label{dfgfihfguhfuhgd}
\mathcal{L}_{0} = -\frac{1}{4}(F_{0}^{\mu\nu})^{2} + \overline{\psi}_{0}(i\slashed{\partial} - m_{0})\psi_{0} - e_{0}\overline{\psi}_{0}\gamma^{\mu}\psi_{0}A_{\mu,\, 0}, 
\end{eqnarray} 
where $\psi_{0}$ and $A_{\mu,\, 0}$ are the bare (the subscript $0$ indicates that the referred quantity is bare) spinor and vector fields. We can study non-ideal quantum electrodynamic effects by defining the $S_{a}$-matrix for non-ideal quantum electrodynamics (NIQED) as 
\begin{eqnarray}\label{dffgfgjisd}
S_{fi,\, a} = \lim_{\substack{t_{1}\rightarrow -\infty \\  t_{2}\rightarrow +\infty}}\bra f T~ e_{a}^{-i\int_{t_{1}}^{t_{2}} dt^{\prime} H_{int}(t^{\prime})} \ket i, 
\end{eqnarray}
where
\begin{eqnarray}\label{efuhgd}
H_{int} = \int d^{3}x ~ e\overline{\psi}\gamma_{\mu}\psi A^{\mu}
\end{eqnarray} 
and the $a$-exponential employed is the non-ideal one, namely $a e_{a}^{x} = e^{ax} - 1 + a$ ($0 < a < 2$) \cite{ArxivNISFT}. 

\section{LO approximation}

\par The first terms to the expansion of both momentum space free electron and free photon propagators and the vertex are, respectively,
 \begin{eqnarray}\label{adfgt}
iD_{\mu\nu,\, a}(p) = \frac{-ig_{\mu\nu}}{p^{2} + i\varepsilon} = \parbox{20mm}{\includegraphics[scale=.7]{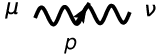}},
\end{eqnarray}
\begin{eqnarray}\label{egfgu}
iS_{F,\, a}(p) = \frac{i}{\slashed{p} - m + i\varepsilon} = ~\parbox{14mm}{\includegraphics[scale=.7]{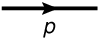}},
\end{eqnarray}
\begin{eqnarray}\label{hjhu}
-ie\Gamma_{\mu,\, a}(p^{\prime}, p, P) = \hspace{1mm} -ie\gamma_{\mu} \hspace{1mm} = \hspace{1mm} ~\parbox{24mm}{\includegraphics[scale=.6]{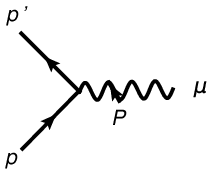}}. 
\end{eqnarray}
Some elementary processes start from first order and others at second order as their LO terms. Now we study LO non-ideal effects by computing the $S_{a}$-matrix perturbatively.

\subsection{Coulomb scattering of electrons}

\par We can compute the differential cross section for the LO non-ideal scattering amplitude of some relativistic electron interacting with a fixed Coulomb potential generated by a static point charge $-Ze$  
\begin{eqnarray}\label{gdfdhjdfhjd}
i\mathcal{M}_{a} = \hspace{1mm} \parbox{24mm}{\includegraphics[scale=.7]{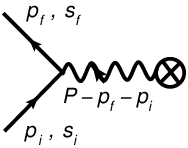}}.  
\end{eqnarray}
The result is the non-ideal Mott formula and its nonrelativistic limit, \emph{i. e.}, the non-ideal Rutherford formula for small velocities $\beta \equiv v/c \ll 1$. Both are given, respectively, by
\begin{eqnarray}\label{adfdhjdfhjd}
\frac{d\sigma_{a}}{d\Omega} = \frac{a^{2}Z^{2}\alpha^{2}\left(1 - \beta^{2}\sin^{2}\dfrac{\theta}{2}\right)}{4\beta^{2}p^{2}\sin^{4}\dfrac{\theta}{2}}\approx \frac{a^{2}Z^{2}\alpha^{2}}{4\beta^{2}p^{2}\sin^{4}\dfrac{\theta}{2}},
\end{eqnarray}

\subsection{Coulomb scattering of positrons}

\par Consider a similar scattering process analogous to that of earlier Subsec., but now for positions instead of electrons. The incoming positron of momentum $\vec{p_{i}}$ and spin $s_{i}$ is described by an outgoing electron with negative energy, momentum $-\vec{p_{i}}$ and spin $-s_{i}$. The outgoing positron is described similarly. The corresponding Feynman diagram is given by 
\begin{eqnarray}\label{jdfhjd}
i\mathcal{M}_{a} = \hspace{1mm} \parbox{24mm}{\includegraphics[scale=.7]{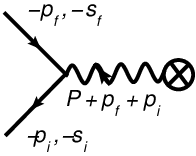}}.  
\end{eqnarray}
The result for the non-ideal Mott formula is the same as that of the earlier Subsec., since these formulae are even functions of $e$. Then they are the same, at least at three-level \cite{GreinerQuantumElectrodynamics}. 

\par But at the next order we have to consider the Feynman diagrams shown just below
%
\begin{eqnarray}\label{fgtjdfhjd}
i\mathcal{M}_{a} = \hspace{1mm} \parbox{22mm}{\includegraphics[scale=.9]{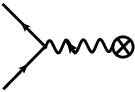}} + \hspace{1mm} \parbox{26mm}{\includegraphics[scale=.7]{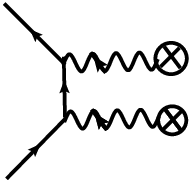}},
\end{eqnarray}
leading to the following non-ideal differential cross section 
\begin{eqnarray}\label{fgtjdfhjdl}
\frac{d\sigma_{a^{\mp}}}{d\Omega} =   \dfrac{a^{2}Z^{2}\alpha^{2}\left[1 - \beta^{2}\sin^{2}\dfrac{\theta}{2} \pm a^{2}\pi Z\alpha\beta\sin\dfrac{\theta}{2}\left(1 - \sin\dfrac{\theta}{2}\right)\right]}{4\beta^{2}p^{2}\sin^{4}\dfrac{\theta}{2}},
\end{eqnarray}
where the minus (plus) sing is for $e^{-}e^{-}$ ($e^{+}e^{+}$) scattering.

\subsection{$e^{-}e^{-}$ scattering}

\par The non-ideal electron-electron scattering amplitude at LO is given by 
\begin{eqnarray}\label{dfdhjdfhjd}
i\mathcal{M}_{a} = \hspace{1mm} \parbox{30mm}{\includegraphics[scale=.6]{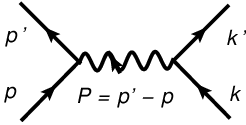}} . 
\end{eqnarray}
It leads, in the nonrelativistic limit \cite{Peskin}, to the repulsive non-ideal Coulomb potential 
\begin{eqnarray}\label{cdfdhjdfghjd}
eV_{a}(r) = a\frac{\alpha}{r},   
\end{eqnarray}
where $\alpha = e^{2}\approx 1/137$ is the fine-structure constant in Gaussian units (and $\hslash = c = 1$) \cite{GreinerQuantumElectrodynamics}. For both non-ideal electron-positron and positron-positron scattering, the expressions for the corresponding non-ideal potential energies have the same absolute values of Eq. (\ref{cdfdhjdfghjd}), but with opposite and same signs, respectively. 

\subsection{$e^{-}p^{+}$ scattering}

\par For the LO non-ideal electron-proton scattering process we have
\begin{eqnarray}\label{dfdfffd}
i\mathcal{M}_{a} = \hspace{1mm} \parbox{30mm}{\includegraphics[scale=.6]{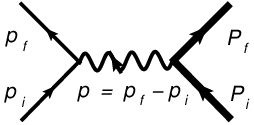}}.
\end{eqnarray}
We can compute the corresponding non-ideal Coulomb potential energy between the electron and proton. The thin line represents the electron, the thicker one symbolizes the proton and the wavy line is associated to the photon exchanged in the process. Here, the proton is approximated by a point-like particle. By applying the electric form factor of the proton at zero momentum transfer, namely $F_{1}(0) = 1$, we obtain
\begin{eqnarray}\label{fghufghj}
eV_{a}(r) = -a\frac{\alpha}{r}.
\end{eqnarray}
Now by considering the internal structure and magnetic moment of the proton  
\begin{eqnarray}\label{adfdfffd}
i\mathcal{M}_{a} = \hspace{1mm} \parbox{30mm}{\includegraphics[scale=.6]{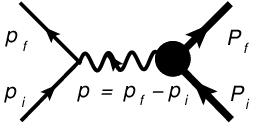}}
\end{eqnarray}
we obtain the non-ideal Rosenbluth's formula
\begin{eqnarray}\label{dfdh}
\frac{d\sigma_{a}}{d\Omega} =  \frac{a^{2}\alpha^{2}\left[ \left(F_{1}^{2} -\dfrac{p^{2}}{4M_{p}^{2}}F_{2}^{2}\right)\cos^{2}\dfrac{\theta}{2} - (F_{1} + F_{2})^{2}\dfrac{p^{2}}{2M_{p}^{2}}\sin^{2}\dfrac{\theta}{2} \right]}{4E^{2}\sin^{4}\dfrac{\theta}{2}\left(1 + \dfrac{2E}{M_{p}}\sin^{2}\dfrac{\theta}{2}\right)},\nonumber \\&&
\end{eqnarray}
where the electron energy is $E$ and $M_{p}$ is the mass of the proton. The factors $F_{1}$ and $F_{2}$ are the electric and magnetic form factors of the proton, respectively.

\subsection{$e^{-}e^{-}$ scattering: Identical particles}

\par We can evaluate the LO non-ideal electron-electron (M$\slashed{o}$ller) scattering amplitude when these particles are \emph{indistinguishable} or \emph{identical} \cite{GreinerQuantumElectrodynamics}. The direct and exchange Feynman graphs contributing to the process at three-level are 
\begin{eqnarray}\label{fdujhdfuhu}
i\mathcal{M}_{a} =  \parbox{34mm}{\includegraphics[scale=.7]{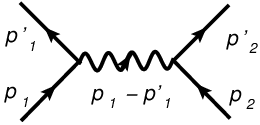}} + ~~ \parbox{30mm}{\includegraphics[scale=.7]{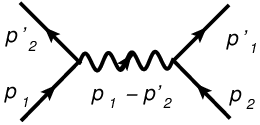}},~~~~~~~~~~~~ 
\end{eqnarray}
whose differential cross section in the ultrarelativistic limit, the non-ideal M$\slashed{o}$ller formula, is given by
\begin{eqnarray}\label{hjdfuhfdh}
\frac{d\sigma_{a}}{d\Omega} =   \frac{a^{2}\alpha^{2}}{8E^{2}}\left( \frac{1 + \cos^{4}\dfrac{\theta}{2}}{\sin^{4}\dfrac{\theta}{2}} + \frac{1 + \sin^{4}\dfrac{\theta}{2}}{\cos^{4}\dfrac{\theta}{2}} - \frac{2}{\sin^{2}\dfrac{\theta}{2}\cos^{2}\dfrac{\theta}{2}} \right).
\end{eqnarray}

\subsection{$e^{-}e^{+}$ scattering: Identical particles}

\par For LO non-ideal electron-positron (Bhabha) scattering, we have to consider both direct and exchange amplitudes \cite{GreinerQuantumElectrodynamics},
\begin{eqnarray}
i\mathcal{M}_{a} = ~~ \parbox{30mm}{\includegraphics[scale=.7]{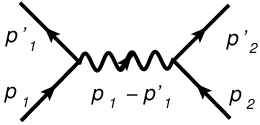}} ~~ + ~~ \parbox{30mm}{\includegraphics[scale=.7]{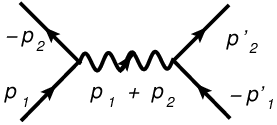}}.
\end{eqnarray}
The corresponding differential non-ideal cross section is given by
\begin{eqnarray}
\frac{d\sigma_{a}}{d\Omega} =   \frac{a^{2}\alpha^{2}}{8E^{2}}\left( \frac{1 + \cos^{4}\dfrac{\theta}{2}}{\sin^{4}\dfrac{\theta}{2}} + \frac{1 + \cos^{2}\theta}{2} - \frac{2\cos^{4}\dfrac{\theta}{2}}{\sin^{2}\dfrac{\theta}{2}} \right).
\end{eqnarray}

\subsection{$e^{+}e^{-}\rightarrow\mu^{+}\mu^{-}\gamma$ scattering}

\par The following Feynman diagram \cite{Schwartz}
\begin{eqnarray}\label{jfufu}
i\mathcal{M}_{a} = \parbox{34mm}{\includegraphics[scale=.7]{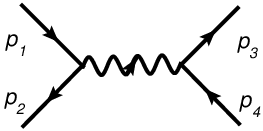}}.
\end{eqnarray}
represents the LO scattering amplitude for the non-ideal process $e^{+}e^{-}\rightarrow\mu^{+}\mu^{-}\gamma$, whose differential cross section can be written as 
\begin{eqnarray}
\dfrac{d\sigma_{a}}{d\Omega} =   \dfrac{a^{2}\alpha^{2}}{4E^{2}}(1 + \cos^{2}\theta).
\end{eqnarray}

\subsection{$e^{-}\gamma$}

\par We can compute the LO non-ideal differential cross section for $e^{-}\gamma$ (Compton) scattering of photons by free electrons \cite{GreinerQuantumElectrodynamics}. Its three-level processes, where the two Feynman diagrams, the direct and exchange, are shown just below
\begin{eqnarray}\label{byughfuhu}
\parbox{35mm}{\includegraphics[scale=.6]{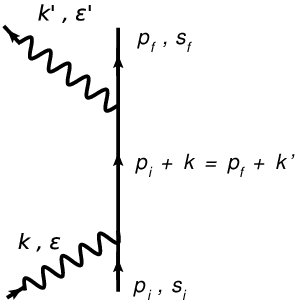}} \hspace{3mm} \parbox{34mm}{\includegraphics[scale=.6]{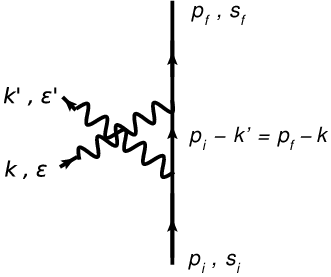}}, \quad
\end{eqnarray}
respectively. Then, the referred differential cross section, non-ideal the Klein-Nishina formula, is given by
\begin{eqnarray}\label{ghuhfdh}
\frac{d\sigma_{a}}{d\Omega} =  \frac{a^{2}\alpha^{2}}{4m^{2}}\left(\frac{\omega^{\prime}}{\omega}\right)^{2}\left[\frac{p_{f}}{p_{i}} + \frac{p_{i}}{p_{f}} + 4(\varepsilon^{\prime}\cdot\varepsilon)^{2} - 2\right],
\end{eqnarray}
where $k\cdot p_{i} = m_{e}\omega$ and $k^{\prime}\cdot p_{i} = m_{e}\omega^{\prime}$.

\subsection{Bremsstrahlung}

\par We can study the LO non-ideal Bremsstrahlung process \cite{GreinerQuantumElectrodynamics}, when a relativistic electron is scattered by some external (static) fixed Coulomb potential. We have to consider the following two Feynman diagrams shown just below
\begin{eqnarray}\label{yughfuhu}
\parbox{20mm}{\includegraphics[scale=.6]{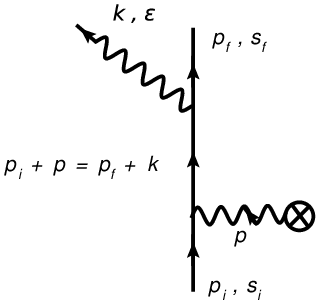}} \hspace{2.0cm} \parbox{20mm}{\includegraphics[scale=.6]{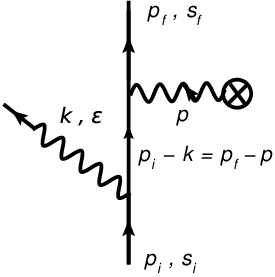}} \quad\quad\quad\quad
\end{eqnarray}
whose differential cross section is given by
\begin{eqnarray}\label{asdfuhjdfuhfdh}
&& d\sigma_{a} = \frac{a^{4}Z^{2}\alpha^{3}}{(2\pi)^{2}}\frac{p_{f}}{p_{i}q^{4}}\frac{d\omega}{\omega}d\Omega_{\gamma}d\Omega_{e} \left[\frac{p_{f}^{2}\sin^{2}\theta_{f}(4E_{i}^{2} - q^{2})}{(E_{f} - p_{f}\cos\theta_{f})^{2}} + \frac{p_{i}^{2}\sin^{2}\theta_{i}(4E_{f}^{2} - q^{2})}{(E_{i} - p_{i}\cos\theta_{i})^{2}} + \right.  \nonumber \\  &&\left.   \right.  \nonumber \\  &&\left. \frac{2\omega^{2}(p_{i}^{2}\sin^{2}\theta_{i} + p_{f}^{2}\sin^{2}\theta_{f})}{(E_{i} - p_{i}\cos\theta_{i})(E_{f} - p_{f}\cos\theta_{f})} -  \frac{2p_{i}p_{f}\sin\theta_{i}\sin\theta_{f}\cos\varphi}{(E_{i} - p_{i}\cos\theta_{i})(E_{f} - p_{f}\cos\theta_{f})(2E_{i}^{2} + 2E_{f}^{2} - q^{2})}\right].~~~~~~~~
\end{eqnarray}


\section{NLO radiative corrections}

\par Now we illustrate how to compute non-ideal effects beyond LO for NIQED.

\subsection{Vacuum polarization contribution to Lamb shift}

\par We can obtain the effect of non-ideal vacuum polarization (NIVP) up to NLO in Eq. (\ref{gdfdhjdfhjd}), where both three-level and one-loop order contributions are displayed just below
\begin{eqnarray}\label{dfffd}
\parbox{22mm}{\includegraphics[scale=.9]{Motthigher2.eps}}  +  \parbox{42mm}{\includegraphics[scale=.9]{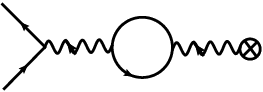}},
\end{eqnarray}
leading, for hydrogen-like atoms, to
\begin{eqnarray}\label{oiuudfghjd}
eV_{a}(r) = -a\left[\frac{Z\alpha}{r} + \frac{4a^{2}Z\alpha^{2}}{15m_{e}^{2}}\delta^{(3)}(\vec{x})\right],   
\end{eqnarray}
where $\delta^{(3)}(\vec{x})$ is the tree-dimensional Dirac delta distribution. The last term of Eq. (\ref{oiuudfghjd}), a result of NIVP, shifts the energy levels of the atom. It is responsible for a small part of the non-ideal Lamb shift. In first-order perturbation theory, the shift from NIVP is given by
\begin{eqnarray}\label{jgfjkljdf}
\Delta E_{nl,\, a}^{NIVP} = - \frac{4a^{3}m_{e}Z^{4}\alpha^{5}}{15\pi n^{3}}\delta_{l0}.
\end{eqnarray}

\subsection{Uehling potential}

\par We can also compute the NLO potential generated by some external point charge -$Ze$. It is named the non-ideal Uehling potential and is given by
\begin{eqnarray}\label{wijikgm}
V_{a}(r) =   -\frac{aZe}{r}\left[1 + \frac{2a^{2}\alpha}{3\pi}\int_{1}^{\infty}d\zeta\left(1 + \frac{1}{2\zeta^{2}}\right)\frac{\sqrt{\zeta^{2} - 1}}{\zeta^{2}}e^{-2m_{e}\zeta r}\right], ~~~~~~~~ 
\end{eqnarray}
whose limits $m_{e}r\ll 1$ and $m_{e}r\gg 1$ furnish the following results
\begin{eqnarray}\label{hrijikgm}
V_{a}(r) \simeq -\frac{aZe}{r} \left[1 + \frac{2a^{2}\alpha}{3\pi} \left(\ln \frac{1}{m_{e}r}  -\frac{5}{6} - C\right)\right],
\end{eqnarray}
\begin{eqnarray}\label{rijikgm}
V_{a}(r) \simeq -\frac{aZe}{r} \left[1 + \frac{a^{2}\alpha}{4\sqrt{\pi}}\frac{e^{-2m_{e}r}}{(m_{e}r)^{3/2}}\right],
\end{eqnarray}
respectively, where $C = 0,5772...$ is the Euler constant. 


\subsection{Anomalous magnetic moment}

\par We can compute the NLO pure-QED non-ideal Landé's $g_{q}$ factor, that is related to the anomalous magnetic moment of the electron. It is due to the non-ideal interaction energy of the electron with some static external electromagnetic field up to NLO, where both three-level and one-loop order contributions are displayed just below 
\begin{eqnarray}\label{cbrhjhua}
\parbox{26mm}{\includegraphics[scale=1]{Motthigher2.eps}}  +  \parbox{24mm}{\includegraphics[scale=1]{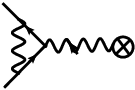}},
\end{eqnarray}
whose value is given by
\begin{eqnarray}\label{adfikgm}
g_{a} = 2\left(1 + \frac{a^{2}}{2}\frac{\alpha}{\pi}\right).
\end{eqnarray}


%

\subsection{$e^{+}e^{-}\rightarrow\mu^{+}\mu^{-}\gamma$ process}

\par We have now to consider the $e^{+}e^{-}\rightarrow\mu^{+}\mu^{-}\gamma$ process, whose total cross section up to NLO is obtained by considering the diagram of Eq. (\ref{jfufu}) and the ones
\begin{eqnarray}
\parbox{34mm}{\includegraphics[scale=.7]{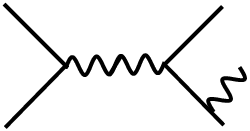}},\hspace{.6cm}  \parbox{26mm}{\includegraphics[scale=.7]{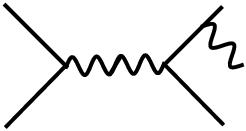}}
\end{eqnarray}
and
\begin{eqnarray}
\parbox{26mm}{\includegraphics[scale=.7]{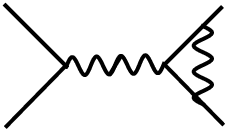}}
\end{eqnarray}
in the following Eq. \cite{Schwartz}
\begin{eqnarray}
2\Re\left(~~\parbox{26mm}{\includegraphics[scale=.7]{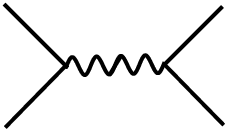}}~~\times~~\parbox{26mm}{\includegraphics[scale=.7]{em3.eps}}~~\right) +  \left(~\parbox{30mm}{\includegraphics[scale=.7]{em1.eps}}~~+~~\parbox{26mm}{\includegraphics[scale=.7]{em2.eps}}~~~\right)^{2}
\end{eqnarray}
that furnishes the result
\begin{eqnarray}\label{uhduhfu}
\sigma_{\text{tot},\, a} = \sigma_{0,\, a}\left(1 + \dfrac{3a^{2}\alpha}{16\pi^{2}}\right),
\end{eqnarray}
where the infrared divergences have been canceled out and
\begin{eqnarray}
\sigma_{0,\, a} = \dfrac{a^{2}\alpha^{2}}{12\pi E^{2}} 
\end{eqnarray}
is the non-ideal total cross section for the process aforementioned at LO. Such a canceling is not possible for non-renormalizable theories defined earlier in the literature and they can not furnish this NLO result. They can furnish only a three-level result as shown for some effective theories in Table \ref{TCSNRT}), thus characterizing them all, at most, as effective ones.


%
%

\subsection{Renormalization group}

\par The corresponding non-ideal Callan-Symanzik equation for the connected ($n,m$)-point function $G_{c,\hspace{.5mm}q}^{(n,m)}(\{x_{i}\};M,e)$ is given by
\begin{eqnarray}\label{abrhjhu}
\left[M\frac{\partial}{\partial M} + \beta_{a}(e)\frac{\partial}{\partial e} + n\gamma_{2,\, a}(e) + m\gamma_{3,\, a}(e) + \gamma_{m_{e},\, a}m_{e}\frac{\partial}{\partial m_{e}}\right]  \times G_{c,\, a}^{(n,m)}(\{x_{i}\};M,e)=0,
\end{eqnarray}
where $n$ and $m$ are the number of fermion and gauge-boson fields, respectively, and $M$ is some arbitrary mass scale \cite{Peskin} and functions $\gamma_{2,\, a}$ and $\gamma_{3,\, a}$ are the corresponding rescaling functions. By applying the non-ideal one-loop values of photon self-energy, electron self-energy and vertex diagrams, respectively
\begin{eqnarray}
\parbox{40mm}{\includegraphics[scale=.6]{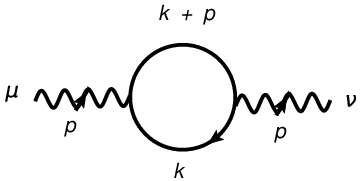}},
\end{eqnarray}
\begin{eqnarray}
\parbox{32mm}{\includegraphics[scale=.7]{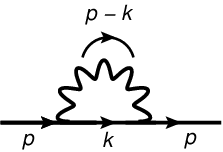}}, 
\end{eqnarray}
\begin{eqnarray}
\parbox{36mm}{\includegraphics[scale=.7]{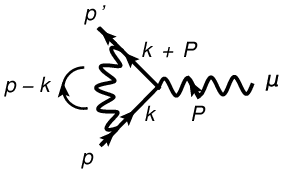}},
\end{eqnarray}
in dimensional regularization in dimension $d = 4$, we have
\begin{eqnarray}
\beta_{a}(e) = -\frac{a^{2}e^{3}}{12\pi^{2}},
\end{eqnarray}
\begin{eqnarray}
\gamma_{2,\, a}(e) = \frac{a^{2}e^{2}}{16\pi^{2}}, 
\end{eqnarray}
\begin{eqnarray}
\gamma_{3,\, a}(e) = \frac{a^{2}e^{2}}{12\pi^{2}},
\end{eqnarray}
\begin{eqnarray}
\gamma_{m,\, a}(e) = \frac{3a^{2}e^{2}}{8\pi^{2}}.
\end{eqnarray}

\subsection{Running coupling constant}

\par In the limit of small distances or $-p^{2} \gg m_{e}^{2}$, we have that the effective non-ideal coupling constant is given by
\begin{eqnarray}\label{ergkgm}
\alpha_{eff,\, a}(p^{2}) = \dfrac{\alpha}{1 - \dfrac{a^{2}\alpha}{3\pi}\log\left(\dfrac{-p^{2}}{Am_{e}^{2}}\right)},
\end{eqnarray}
where $A = \exp(5/3)$.

\section{Interpretation of the results}

\par All the effects approached in this Letter present non-ideal contributions, \emph{i. e.}, depend on the non-ideal parameter $a$ and display stronger (weaker) interactions for $a > 1$ ($a < 1$). Clear examples are the cross sections that assume higher (lower) values for $a > 1$ ($a < 1$). This interpretations is in agreement with that for systems undergoing continuous phase transitions in low energies \cite{ArxivNISFT}.  

\par Another important point to mention is that earlier similar but non-renormalizable theories defined recently \cite{ArxivNISFT} can not describe non-ideal physical effects up to NLO but are limited to furnish results only at LO, thus characterizing them all, at most, as effective ones. One example is the $e^{+}e^{-}\rightarrow\mu^{+}\mu^{-}\gamma$ process, whose non-ideal total cross sections at LO (not up to NLO as in Eq. (\ref{uhduhfu}) for NIQED but just the LO term of that Eq.) are displayed in Table \ref{TCSNRT} for some others non-renormalizable theories. 
\begin{table}[t]
\caption{Total cross sections for some non-renormalizable theories at LO.}
\begin{tabular}{p{8cm}}
\hline
[theory] \vspace{.1mm} $\sigma_{\text{tot}}$    \\
\hline
\hspace{.1mm} \\
\cite{CARVALHO2024139487} \vspace{.1mm}  $\sigma_{\text{tot},\, \delta_{KLS}} \equiv \sigma_{0,\, \delta_{KLS}} = \dfrac{(1 - 2\delta_{KLS})^{2}\alpha^{2}}{12\pi E^{2}}$  \\
\hspace{.1mm} \\
\cite{CARVALHO2023138187}   \vspace{.1mm} $\sigma_{\text{tot},\, \delta_{KLS}} \equiv \sigma_{0,\, \delta_{KLS}} = \dfrac{(1 - \gamma_{KLS})^{2}\alpha^{2}}{12\pi E^{2}}$   \\
\hspace{.1mm} \\
\cite{ALVES2023138005}   \vspace{.1mm} $\sigma_{\text{tot},\, \kappa} \equiv \sigma_{0,\, \delta_{KLS}} = \dfrac{(1 - 2\delta_{KLS})^{2}\alpha^{2}}{12\pi E^{2}}$  \\
\hspace{.1mm} \\
\cite{CARVALHO2023137683}   \vspace{.1mm} $\sigma_{\text{tot},\, q} \equiv \sigma_{0,\, q} = \dfrac{q^{2}\alpha^{2}}{12\pi E^{2}}$ \\
\hspace{.1mm} \\
\cite{Shafee}   \vspace{.1mm} $\sigma_{\text{tot},\, q} \equiv \sigma_{0,\, q} = \dfrac{(3 - 2q)^{2}\alpha^{2}}{12\pi E^{2}}$  \\
\hspace{.1mm} \\
\cite{Fibonacci}  \vspace{.1mm} $\sigma_{\text{tot},\, q_{1},\, q_{2}} \equiv \sigma_{0,\, q_{1},\, q_{2}} = \dfrac{(q_{1}^{2} + q_{1}^{2})^{2}\alpha^{2}}{48\pi E^{2}}$  \\
\hspace{.1mm} \\
\cite{Jackson}   \vspace{.1mm} $\sigma_{\text{tot},\, q} \equiv \sigma_{0,\, q} = \dfrac{(1 + q)^{2}\alpha^{2}}{48\pi E^{2}}$  \\
\hspace{.1mm} \\
\hline
 \end{tabular}
\label{TCSNRT}
\end{table}

\section{Conclusions}

\par In this Letter we have introduced a field-theoretic formulation for describing non-ideal quantum electrodynamic effects. It generalizes its ideal counterpart and is valid in the non-ideal domain, \emph{i. e.}, it depends on the non-ideal parameter $a$ and displays stronger (weaker) interactions for $a > 1$ ($a < 1$). Clear examples are the cross sections that assume higher (lower) values for $a > 1$ ($a < 1$). This interpretations is in agreement with that for systems undergoing continuous phase transitions in low energies \cite{ArxivNISFT}. We have computed some non-ideal elementary processes both at LO as well as up to NLO. Furthermore we presented some physical effect, namely the $e^{+}e^{-}\rightarrow\mu^{+}\mu^{-}\gamma$ process, taken up to NLO which is not described by earlier defined similar but non-renormalizable theories up to NLO but just at LO thus characterizing them all, at most, as effective ones. We recover all the corresponding ideal results when we take the limit $a\rightarrow 1$, where $a$ is a parameter that encodes non-ideal effects. All the results show an interplay between non-ideal properties and fluctuations.

\section*{Declaration of competing interest}

\par The authors declare that they have no known competing financial interests or personal relationships that could have appeared to influence the work reported in this paper.

\section*{Acknowledgments}

\par PRSC would like to thank the Brazilian funding agencies CAPES and CNPq (Grant: Produtividade 306130/2022-0) for financial support.

\bibliography{apstemplate}

\providecommand{\noopsort}[1]{}\providecommand{\singleletter}[1]{#1}%
\begin{thebibliography}{13}%
\makeatletter
\providecommand \@ifxundefined [1]{%
 \@ifx{#1\undefined}
}%
\providecommand \@ifnum [1]{%
 \ifnum #1\expandafter \@firstoftwo
 \else \expandafter \@secondoftwo
 \fi
}%
\providecommand \@ifx [1]{%
 \ifx #1\expandafter \@firstoftwo
 \else \expandafter \@secondoftwo
 \fi
}%
\providecommand \natexlab [1]{#1}%
\providecommand \enquote  [1]{``#1''}%
\providecommand \bibnamefont  [1]{#1}%
\providecommand \bibfnamefont [1]{#1}%
\providecommand \citenamefont [1]{#1}%
\providecommand \href@noop [0]{\@secondoftwo}%
\providecommand \href [0]{\begingroup \@sanitize@url \@href}%
\providecommand \@href[1]{\@@startlink{#1}\@@href}%
\providecommand \@@href[1]{\endgroup#1\@@endlink}%
\providecommand \@sanitize@url [0]{\catcode `\\12\catcode `\$12\catcode
  `\&12\catcode `\#12\catcode `\^12\catcode `\_12\catcode `\%12\relax}%
\providecommand \@@startlink[1]{}%
\providecommand \@@endlink[0]{}%
\providecommand \url  [0]{\begingroup\@sanitize@url \@url }%
\providecommand \@url [1]{\endgroup\@href {#1}{\urlprefix }}%
\providecommand \urlprefix  [0]{URL }%
\providecommand \Eprint [0]{\href }%
\providecommand \doibase [0]{http://dx.doi.org/}%
\providecommand \selectlanguage [0]{\@gobble}%
\providecommand \bibinfo  [0]{\@secondoftwo}%
\providecommand \bibfield  [0]{\@secondoftwo}%
\providecommand \translation [1]{[#1]}%
\providecommand \BibitemOpen [0]{}%
\providecommand \bibitemStop [0]{}%
\providecommand \bibitemNoStop [0]{.\EOS\space}%
\providecommand \EOS [0]{\spacefactor3000\relax}%
\providecommand \BibitemShut  [1]{\csname bibitem#1\endcsname}%
\let\auto@bib@innerbib\@empty
\bibitem [{\citenamefont {Carvalho}(2025)}]{ArxivNISFT}%
  \BibitemOpen
  \bibfield  {author} {\bibinfo {author} {\bibfnamefont {P.~R.~S.}\
  \bibnamefont {Carvalho}},\ }\href@noop {} {\emph {\bibinfo {title}
  {arXiv:2506.21424}}}\ (\bibinfo {year} {2025})\BibitemShut {NoStop}%
\bibitem [{\citenamefont {Wilson}\ and\ \citenamefont
  {Kogut}(1974)}]{Wilson197475}%
  \BibitemOpen
  \bibfield  {author} {\bibinfo {author} {\bibfnamefont {K.~G.}\ \bibnamefont
  {Wilson}}\ and\ \bibinfo {author} {\bibfnamefont {J.}~\bibnamefont {Kogut}},\
  }\href@noop {} {\bibfield  {journal} {\bibinfo  {journal} {Phys. Rep.}\
  }\textbf {\bibinfo {volume} {12}},\ \bibinfo {pages} {75} (\bibinfo {year}
  {1974})}\BibitemShut {NoStop}%
\bibitem [{\citenamefont {Wilson}(1979)}]{WilsonSciAme}%
  \BibitemOpen
  \bibfield  {author} {\bibinfo {author} {\bibfnamefont {K.~G.}\ \bibnamefont
  {Wilson}},\ }\href@noop {} {\bibfield  {journal} {\bibinfo  {journal} {Sci.
  Am.}\ }\textbf {\bibinfo {volume} {241}},\ \bibinfo {pages} {158} (\bibinfo
  {year} {1979})}\BibitemShut {NoStop}%
\bibitem [{\citenamefont {Schwartz}(2013)}]{Schwartz}%
  \BibitemOpen
  \bibfield  {author} {\bibinfo {author} {\bibfnamefont {M.~D.}\ \bibnamefont
  {Schwartz}},\ }\href@noop {} {\emph {\bibinfo {title} {Quantum Field Theory
  and the Standard Model}}}\ (\bibinfo  {publisher} {Cambridge University
  Press},\ \bibinfo {year} {2013})\BibitemShut {NoStop}%
\bibitem [{\citenamefont {Peskin}\ and\ \citenamefont
  {Schroeder}(1995)}]{Peskin}%
  \BibitemOpen
  \bibfield  {author} {\bibinfo {author} {\bibfnamefont {M.~E.}\ \bibnamefont
  {Peskin}}\ and\ \bibinfo {author} {\bibfnamefont {D.~V.}\ \bibnamefont
  {Schroeder}},\ }\href@noop {} {\emph {\bibinfo {title} {An Introduction to
  Quantum Field Theory}}}\ (\bibinfo  {publisher} {Westview Press, New York},\
  \bibinfo {year} {1995})\BibitemShut {NoStop}%
\bibitem [{\citenamefont {Greiner}\ and\ \citenamefont
  {Reinhardt}(2009)}]{GreinerQuantumElectrodynamics}%
  \BibitemOpen
  \bibfield  {author} {\bibinfo {author} {\bibfnamefont {W.}~\bibnamefont
  {Greiner}}\ and\ \bibinfo {author} {\bibfnamefont {J.}~\bibnamefont
  {Reinhardt}},\ }\href@noop {} {\emph {\bibinfo {title} {Quantum
  Electrodynamics}}},\ \bibinfo {edition} {4th}\ ed.\ (\bibinfo {year}
  {Springer, 2009})\BibitemShut {NoStop}%
\bibitem [{\citenamefont {Carvalho}(2024)}]{CARVALHO2024139487}%
  \BibitemOpen
  \bibfield  {author} {\bibinfo {author} {\bibfnamefont {P.~R.~S.}\
  \bibnamefont {Carvalho}},\ }\href@noop {} {\bibfield  {journal} {\bibinfo
  {journal} {Eur. Phys. J. Plus}\ }\textbf {\bibinfo {volume} {139}},\ \bibinfo
  {pages} {487} (\bibinfo {year} {2024})}\BibitemShut {NoStop}%
\bibitem [{\citenamefont {Carvalho}(2023{\natexlab{a}})}]{CARVALHO2023138187}%
  \BibitemOpen
  \bibfield  {author} {\bibinfo {author} {\bibfnamefont {P.~R.~S.}\
  \bibnamefont {Carvalho}},\ }\href@noop {} {\bibfield  {journal} {\bibinfo
  {journal} {Phys. Lett. B}\ }\textbf {\bibinfo {volume} {846}},\ \bibinfo
  {pages} {138187} (\bibinfo {year} {2023}{\natexlab{a}})}\BibitemShut
  {NoStop}%
\bibitem [{\citenamefont {Alves}\ \emph {et~al.}(2023)\citenamefont {Alves},
  \citenamefont {Neto}, \citenamefont {Lima}, \citenamefont {Alves},\ and\
  \citenamefont {Carvalho}}]{ALVES2023138005}%
  \BibitemOpen
  \bibfield  {author} {\bibinfo {author} {\bibfnamefont {T.~F.~A.}\
  \bibnamefont {Alves}}, \bibinfo {author} {\bibfnamefont {J.~F.~S.}\
  \bibnamefont {Neto}}, \bibinfo {author} {\bibfnamefont {F.~W.~S.}\
  \bibnamefont {Lima}}, \bibinfo {author} {\bibfnamefont {G.~A.}\ \bibnamefont
  {Alves}}, \ and\ \bibinfo {author} {\bibfnamefont {P.~R.~S.}\ \bibnamefont
  {Carvalho}},\ }\href@noop {} {\bibfield  {journal} {\bibinfo  {journal}
  {Phys. Lett. B}\ }\textbf {\bibinfo {volume} {843}},\ \bibinfo {pages}
  {138005} (\bibinfo {year} {2023})}\BibitemShut {NoStop}%
\bibitem [{\citenamefont {Carvalho}(2023{\natexlab{b}})}]{CARVALHO2023137683}%
  \BibitemOpen
  \bibfield  {author} {\bibinfo {author} {\bibfnamefont {P.~R.~S.}\
  \bibnamefont {Carvalho}},\ }\href {\doibase
  https://doi.org/10.1016/j.physletb.2023.137683} {\bibfield  {journal}
  {\bibinfo  {journal} {Phys. Lett. B}\ }\textbf {\bibinfo {volume} {838}},\
  \bibinfo {pages} {137683} (\bibinfo {year} {2023}{\natexlab{b}})}\BibitemShut
  {NoStop}%
\bibitem [{\citenamefont {Shafee}(2007)}]{Shafee}%
  \BibitemOpen
  \bibfield  {author} {\bibinfo {author} {\bibfnamefont {F.}~\bibnamefont
  {Shafee}},\ }\href@noop {} {\bibfield  {journal} {\bibinfo  {journal} {IMA J.
  Appl. Math.}\ }\textbf {\bibinfo {volume} {72}},\ \bibinfo {pages} {785}
  (\bibinfo {year} {2007})}\BibitemShut {NoStop}%
\bibitem [{\citenamefont {Arik}\ \emph {et~al.}(1992)\citenamefont {Arik},
  \citenamefont {Demircan}, \citenamefont {Turgut}, \citenamefont {Ekinci},\
  and\ \citenamefont {Mungan}}]{Fibonacci}%
  \BibitemOpen
  \bibfield  {author} {\bibinfo {author} {\bibfnamefont {M.}~\bibnamefont
  {Arik}}, \bibinfo {author} {\bibfnamefont {E.}~\bibnamefont {Demircan}},
  \bibinfo {author} {\bibfnamefont {T.}~\bibnamefont {Turgut}}, \bibinfo
  {author} {\bibfnamefont {L.}~\bibnamefont {Ekinci}}, \ and\ \bibinfo {author}
  {\bibfnamefont {M.}~\bibnamefont {Mungan}},\ }\href@noop {} {\bibfield
  {journal} {\bibinfo  {journal} {Z. Phys. C}\ }\textbf {\bibinfo {volume}
  {55}},\ \bibinfo {pages} {89} (\bibinfo {year} {1992})}\BibitemShut {NoStop}%
\bibitem [{\citenamefont {Jackson}(1908)}]{Jackson}%
  \BibitemOpen
  \bibfield  {author} {\bibinfo {author} {\bibfnamefont {F.}~\bibnamefont
  {Jackson}},\ }\href@noop {} {\bibfield  {journal} {\bibinfo  {journal}
  {Trans. R. Soc. Edin.}\ }\textbf {\bibinfo {volume} {46}},\ \bibinfo {pages}
  {253} (\bibinfo {year} {1908})}\BibitemShut {NoStop}%
\end{thebibliography}%

\end{document}